\documentclass[preprint,journal]{vgtc}       

\ifpdf
  \pdfoutput=1\relax                   
  \usepackage{graphicx}                
  \DeclareGraphicsExtensions{.pdf,.png,.jpg,.jpeg} 
\else
  \usepackage{graphicx}                
  \DeclareGraphicsExtensions{.eps}     
\fi%

\graphicspath{{figures/}{pictures/}{images/}{./}} 

\usepackage{microtype}                 
\PassOptionsToPackage{warn}{textcomp}  
\usepackage{textcomp}                  
\usepackage{mathptmx}                  
\usepackage{times}                     
\usepackage{cite}                      
\usepackage{tabu}                      
\usepackage{booktabs}                  

\usepackage{tcolorbox}

\definecolor{tagbordercolor}{rgb}{0.8, 0.8, 0.8}
\definecolor{tagbgcolor}{rgb}{0.9, 0.9, 0.9}

\newtcbox{\tag}{nobeforeafter, colframe=tagbordercolor,
colback=tagbgcolor, boxrule=0.5pt, arc=1pt,
  boxsep=0pt,left=2pt,right=2pt,top=1.5pt,bottom=2pt,tcbox raise base}

\definecolor{myPurple}{rgb}{0.62, 0.40, 0.74}
\definecolor{myBlue}{rgb}{0.08, 0.46, 0.70}

\usepackage{color}
\usepackage{enumitem}

\newcommand{\name}[1]{\textsl{\textsc{ActiVis}}}

\preprinttext{}

\onlineid{0}

\vgtccategory{Research}
\vgtcpapertype{}

\title{\name{}: 
Visual Exploration of Industry-Scale \\Deep Neural Network Models}

\author{Minsuk Kahng, Pierre Y. Andrews, Aditya Kalro, and Duen Horng (Polo) Chau}
\authorfooter{
\item
 Minsuk Kahng and Duen Horng (Polo) Chau are with Georgia Institute of Technology. E-mail: \{kahng,polo\}@gatech.edu.
 This work was done while Minsuk Kahng was at Facebook.
\item
 Pierre Y. Andrews and Aditya Kalro are with Facebook. E-mail: \{mortimer,adityakalro\}@fb.com.
 \item
 This paper will be presented at the IEEE Conference on Visual Analytics Science and Technology (VAST) in October 2017 and published in the IEEE Transactions on Visualization and Computer Graphics (TVCG), Vol. 24, No. 1, January 2018.
}

\shortauthortitle{Kahng \MakeLowercase{\textit{et al.}}: \name{}: 
Visual Exploration of Industry-Scale Deep Neural Network Models}

\abstract{
While deep learning models have achieved state-of-the-art accuracies for many prediction tasks, 
understanding these models remains a challenge.
Despite the recent interest in developing visual tools to help users interpret deep learning models, 
the complexity and wide variety of models deployed in industry, and the large-scale datasets that they used, 
pose unique design challenges that are inadequately addressed by existing work. 
Through participatory design sessions with over 15 researchers and engineers at Facebook, we have developed, deployed, and iteratively improved \name{},
an interactive visualization system for interpreting large-scale deep learning models and results. 
By tightly integrating multiple coordinated views, such as a \textit{computation graph} overview of the model architecture, and
a \textit{neuron activation} view for pattern discovery and comparison, 
users can explore complex deep neural network models 
at both the instance- and subset-level.
\name{} has been deployed on Facebook's machine learning platform.
We present case studies with Facebook researchers and engineers, and usage scenarios of how \name{} may work with different models.

} 


\CCScatlist{ 
 \CCScat{H.5.2}{Information Interfaces and Presentation}{User Interfaces}
}

\teaser{
  \centering
  \includegraphics[width=\linewidth]{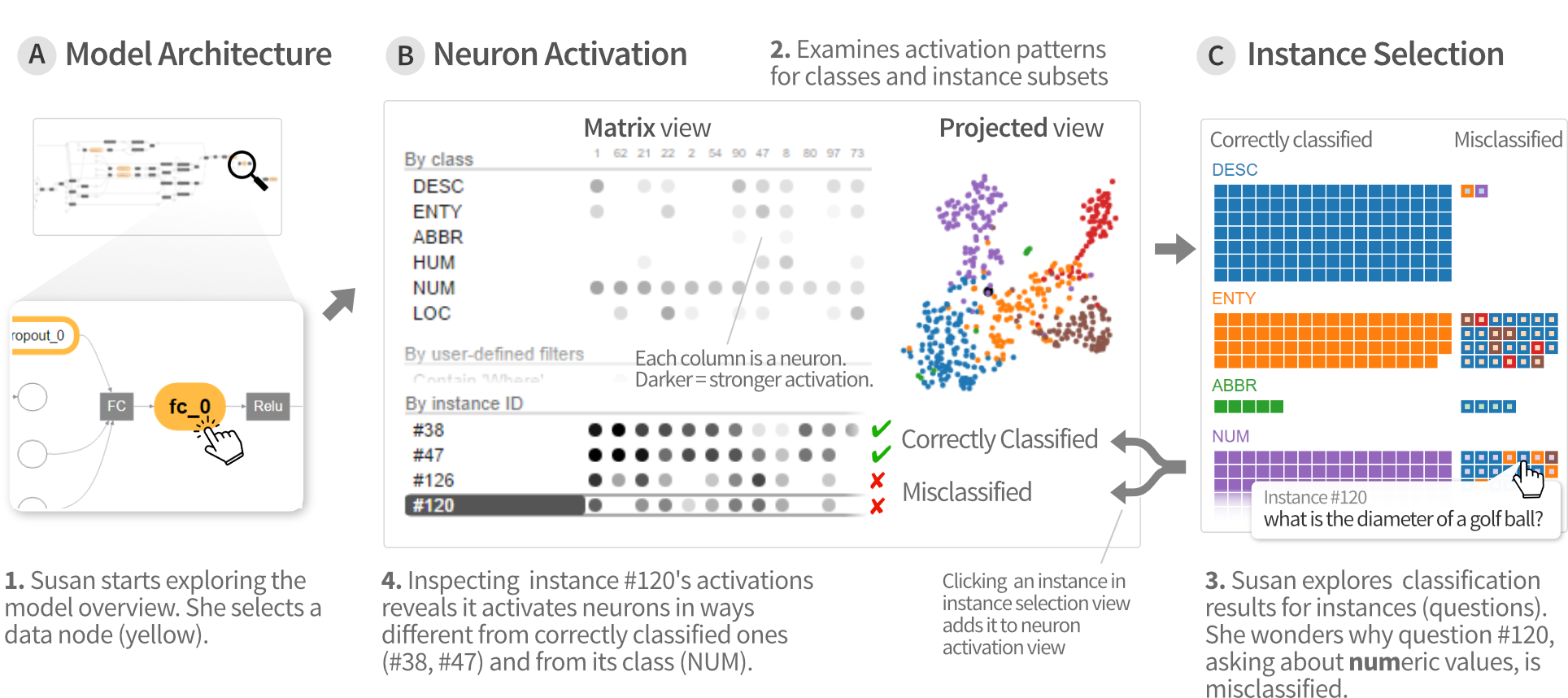}
  \caption{
\name{} integrates several coordinated views to support exploration of complex deep neural network models, at both instance- and subset-level. 
\textbf{1.} Our user Susan starts exploring the model architecture, through its \textit{computation graph} overview (at A). 
Selecting a \textit{data node} (in yellow) displays its \textit{neuron activations} (at B).
\textbf{2.} The \textit{neuron activation matrix view} shows the activations for instances and instance subsets;
the \textit{projected view} displays the 2-D projection of instance activations.
\textbf{3.} From the \textit{instance selection} panel (at C), she explores individual instances and their classification results.
\textbf{4.} Adding instances to the matrix view enables comparison of activation patterns across instances, subsets, and classes, revealing causes for misclassification.
  }
	\label{fig:teaser}
}

\vgtcinsertpkg

\begin{document}

\firstsection{Introduction}

\maketitle

Deep learning has led to major breakthroughs in various domains, such as computer vision, natural language processing, and healthcare. 
Many technology companies, like Facebook, have been increasingly adopting deep learning models for their products~\cite{deeptext,abadi2016tensorflow,covington2016deep}. 
While powerful deep neural network models have significantly improved prediction accuracy, understanding these models remains a challenge. 
Deep learning models are more difficult to interpret than most existing machine learning models, because they capture nonlinear hidden structures of data using a huge number of parameters.
Therefore, in practice, people often use them as ``black boxes'', 
which could be detrimental because when the models do not perform satisfactorily, users would not understand the causes or know how to fix them ~\cite{kulesza2015principles,ribeiro2016why}.

Despite the recent increasing interest in developing visual tools to help users interpret deep learning models~\cite{liu2017towards,yosinski2015understanding,chung2016revacnn,smilkov2016embedding},
the complexity and wide variety of models deployed in industry, and the large-scale datasets that they use, pose unique challenges that are inadequately addressed by existing work.
For example, deep learning tasks in industry often involve  different types of data, 
including text and numerical data; 
however most existing visualization research targets image datasets~\cite{yosinski2015understanding}.
Furthermore, in designing interpretation tools for real-world use and deployment at technology companies,
it is a high priority that the tools be flexible and generalizable to the wide variety of models and datasets that the companies use for their many products and services.
These observations motivate us to design and develop a visualization tool for interpreting industry-scale deep neural network models, one that can work with a wide range of models, and can be readily deployed on  Facebook's machine learning platform.

Through participatory design with researchers, data scientists, and engineers at Facebook,
we have identified common analysis strategies that they use to interpret machine learning models.
Specifically, we learned that both \textbf{instance-} and \textbf{subset-based} exploration approaches are common and effective.
Instance-based exploration (e.g., how individual instances contribute to a model's accuracy) have demonstrated success in a number of machine learning tasks~\cite{kulesza2015principles,amershi2015modeltracker,patel2010gestalt}.
As individual instances are familiar to users, exploring by instances accelerates model understanding.
Another effective strategy is to leverage input features or instance subsets specified by users~\cite{krause2016interacting,kulesza2015principles}.
Slicing results by features helps reveal relationships between data attributes and machine learning algorithms' outputs
~\cite{mcmahan2013ad,kahng2016visual,patel2010gestalt}. 
Subset-based exploration is especially beneficial when dealing with huge datasets in industry, 
which may consist of millions or billions of data points.
Interpreting model results at a higher, more abstract level helps drive down computation time, 
and help user develop general sense about the models.

Our tool, called \textbf{\name{}}, aims to support both interpretation strategies for visualization and comparison of multiple instances and subsets.
\name{} is an interactive visualization system for deep neural network models that 
(1) unifies instance- and subset-level inspections, 
(2) tightly integrates overview of complex models and localized inspection,
and (3) scales to a variety of industry-scale datasets and models.
\name{} visualizes how \textit{neurons} are activated by user-specified instances or instance subsets, 
to help users understand how a model derives its predictions. 
Users can freely define subsets with raw data attributes, transformed features, and output results,
enabling model inspection from multiple angles.
While many existing deep learning visualization tools support  instance-based exploration~\cite{harley2015isvc,yosinski2015understanding,chung2016revacnn,smilkov2016embedding,convnetjs},
\name{} is the first tool that simultaneously supports instance- and subset-based exploration of the deep neural network models.
In addition, to help users get a high-level overview of the model, \name{} provides a graph-based representation of the model architecture, from which the user can drill down to perform localized inspection of activations at each model layer (node).

\textbf{Illustrative scenario.}
To illustrate how \name{} works in practice,
consider our user Susan who is training a word-level \textit{convolutional neural network} (CNN) model~\cite{kim2014convolutional} to classify question sentences into one of six categories  (e.g., whether a question asks about \textit{numeric} values, as in \textit{``what is the diameter of a golf ball?''}).
Her dataset is part of the TREC question answering data collections\footnote{\small \url{http://cogcomp.cs.illinois.edu/Data/QA/QC/}} \cite{li2002learning}.

Susan is new to using this CNN model, 
so she decides to start by using its default training parameters.
After training  completes, she launches \name{}, which runs in a web browser.
\name{} provides an overview of the model by displaying its architecture as a computation graph (\autoref{fig:teaser}A, top),
summarizing the model structure.
By exploring the graph, Susan learns about the kind of operations (e.g., convolution) that are performed, and how they are combined in the model.

Based on her experience working with other deep learning models, 
she knows that a model's performance is strongly correlated with its last hidden layer, thus it would be informative to analyze that layer.
In \name{}, a layer is represented as a rounded rectangular \textit{node} (highlighted in yellow, in \autoref{fig:teaser}A, bottom).

Susan clicks the node for the last hidden layer,
and \name{} displays the layer's \textit{neuron activation} in a panel (\autoref{fig:teaser}B):
the \textit{neuron activation matrix view} on the left shows how neurons (shown as columns) respond to instances from different classes (rows); and 
the \textit{projected view} on the right shows the 2-D projection of instance activations.

In the \textit{matrix view}, stronger neuron activations are shown in darker gray.
Susan sees that the activation patterns for the six classes (rows) are quite visually distinctive, which may indicate satisfactory classification.
However, in the \textit{projected view}, instances from different classes are not clearly separated,
which suggests some degree of misclassification.

To examine the misclassified instances and to investigate why they are mislabeled, 
Susan brings up the \textit{instance selection panel} (\autoref{fig:teaser}C).
The classification results for the \textcolor{myPurple}{\textbf{\sffamily \small NUM}}ber class alarm Susan, as many instances in that class are misclassified (shown in right column).
She examines their associated question text by mouse-overing them, which shows the text in popup tooltips.
She wants to compare the activation patterns of  the correctly classified instances with those of the misclassified.
So she adds two correct instances (\#38, \#47) and two misclassified instances (\#120, \#126) to the \textit{neuron activation matrix view} --- indeed, their activation patterns are very different (\autoref{fig:teaser}.4).

Taking a closer look at the \textit{instance selection panel}, Susan sees that many instances have \textcolor{myBlue}{blue} borders, meaning they are misclassified as \textcolor{myBlue}{\textbf{\sffamily \small DESC}}ription. 
Inspecting the instances' text reveals that they often begin with \textit{``What is''}, which is typical for questions asking for descriptions, though they are also common for other question types,
as in \textit{``What is the diameter of a golf ball?''} which is a numeric question (\autoref{fig:teaser}.3).

To understand the extent to which instances starting with \textit{``What is''} are generally misclassified by the model, Susan creates an \textit{instance subset} for them, and \name{} adds this subset as a new row in the \textit{neuron activation matrix view}.
Susan cannot discern any visual patterns from the subset's seemingly scattered, random neuron activations, suggesting that the model may not yet have learned effective ways to  distinguish between the different intents of \textit{``What is''} questions.
Based on this finding, she proceeds to train more models with different parameters (e.g., consider longer $n$-grams) to better classify these questions.

\name{} integrates multiple coordinated views to enable Susan to work with complex models, and to flexibly explore them at  instance- and subset-level, helping her discover and narrow in to specific issues.

\textbf{Deployment.}
\name{} has been deployed on the machine learning platform at Facebook.
A developer can visualize a  deep learning model  using \name{} by adding only a few lines of code, which instructs the model's training process to generate data needed for \name{}.
\name{} users at Facebook (e.g., data scientists) can then train models and use \name{} via \textit{FBLearner Flow}~\cite{flow,andrews2016productionizing}, Facebook's internal machine learning web interface, without writing any additional code.

\vspace{3pt}
\noindent \name{}'s main contributions include:

\begin{itemize}[topsep=0mm, itemsep=0mm, leftmargin=3mm]

\item A novel visual representation that unifies instance- and subset-level inspections of neuron activations, which facilitates comparison of activation patterns for multiple instances and instance subsets.
Users can flexibly specify subsets using input features, labels, or any intermediate outcomes in a machine learning pipeline (\autoref{sec:main1}).

\item An interface that tightly integrates an overview of graph-structured complex models and local inspection of neuron activations, 
allowing users to explore the model at different levels of abstraction (\autoref{sec:main2}).

\item A deployed system scaling to large datasets and models (\autoref{sec:main3}). 

\item Case studies with Facebook engineers and data scientists that highlight how \name{} helps them with their work, and usage scenarios that describe how \name{} may work with different models (\autoref{sec:evaluations}).

\end{itemize}

\section{Related Work}

\subsection{Machine Learning Interpretation through Visualization}

As the complexity of machine learning algorithms increases, many researchers have recognized the importance of model interpretation and developed interactive tools to help users better understand them~\cite{kulesza2011oriented, gleicher2013explainers,krause2016interacting,van2011baobabview,ribeiro2016why,choo2010ivisclassifier}.
While overall model accuracy can be used to select models, users often want to understand why and when a model would perform better than others, so that they can trust the model and know how to further improve it.
In developing interpretation tools, revealing relationships between data and models is one of the the most important design goals~\cite{patel2008investigating,patel2010gestalt}.
Below we present two important analytics strategies that existing works adopt to help users understand how data respond to machine learning models.

\textbf{Instance-based exploration.}
A widely-used approach to understanding complex algorithms is by tracking how an example (i.e., training or test instance) behaves inside the models.
Kulesza et al.~\cite{kulesza2015principles} presented an interactive system that explains how models made predictions for each instance.
Amershi et al.~\cite{amershi2015modeltracker} developed ModelTracker, a visualization tool that shows the distribution of instance scores for binary classification tasks and allows users to examine each instance individually. The researchers from the same group recently extended their work for multi-classification tasks~\cite{ren2017squares}.
While the above-mentioned tools were designed for model-agnostic, there are also tools designed specifically for neural network  models~\cite{harley2015isvc,convnetjs,smilkov2016directmanipulation}. These tools enable users to pick an instance and feed it to the models and show how the parameters of the models change. We will describe them in more detail shortly, in \autoref{sec:deep}.

\textbf{Feature- and subset-based exploration.}
While instance-based exploration is helpful for tracking how models respond to individual examples, feature- or subset-based exploration enables users to  better understand the relationships between data and models, as machine learning features make it possible for instances to be grouped and sliced in multiple ways.
Researchers have utilized \textit{features} to visually describe how the models captured the structure of datasets~\cite{kulesza2015principles,krause2016interacting, krause2014infuse, brooks2015featureinsight}.
Kulesza et al.~\cite{kulesza2015principles} used the importance weight of each feature in the Naive Bayes algorithm, and Krause et al.~\cite{krause2016interacting} used \textit{partial dependence} to show the relationships between features and results.
To enable users to analyze results not only by predefined features, researchers have developed tools that enable users to specify  instance subsets.
Specifying groups  can be a good first step for analyzing machine learning results~\cite{krause2016supporting}, as it provides users with an effective way for analyzing complex multidimensional data.
In particular, people in the medical domain often perform  similar processes, called \textit{cohort construction}, and Krause et al.~\cite{krause2016supporting} developed an interactive tool that helps this process.
McMahan et al.~\cite{mcmahan2013ad} presented their internal tool that allows users to visually compare the performance differences between models by subsets.
\textit{MLCube}~\cite{kahng2016visual} enabled users to interactively explore and define instance subsets using both raw data attributes and transformed features, and compute evaluation metrics over the subsets.

\subsection{Interactive Visualization of Deep Learning Models}
\label{sec:deep}

Deep  learning  has become very popular, largely thanks to the state-of-the-art performance achieved by convolutional neural network models, commonly used for analyzing image datasets in computer vision.
Since deep neural network models typically consist of many parameters, researchers have recognized deep learning interpretation as an important research area.
A common approach is to show \textit{filters} or \textit{activations} for each neural network \textit{layer}.
This helps users understand what the models have learned in the hidden structure throughout the layers.

\textbf{Interactive visualization tools.}
A number of interactive tools have been developed to effectively visualize the activation information.
Tzeng and Ma~\cite{tzeng2005opening} was one of the first visualization tools designed for neural network models. 
While it did not target deep networks, 
it represented each neuron as a node and visualized a given instance's activations.
This idea has been extended to the case of deep neural networks.
Karpathy~\cite{convnetjs}
visualized the activations for each layer of a neural network on his website.
Harley~\cite{harley2015isvc} developed an interactive prototype that shows activations for a given instance.
Smilkov et al.~\cite{smilkov2016directmanipulation} developed an interactive prototype for educational purposes, called \textit{TensorFlow Playground}, which visualized training parameters to help users  explore how models process a given instance to make predictions.
However, these tools do not scale to large dataset or the complex models commonly used in industry.

\textbf{Towards scalable visualization systems.}
CNNVis~\cite{liu2017towards} is an interactive visual analytics system designed for convolutional networks. It modeled neurons as a directed graph and utilized several techniques to make it scalable. 
For example, it uses hierarchical clustering to group neurons  and uses bi-directional edge bundling to summarize edges among neurons.
They also compute average activations for instances from the same class.
However, users cannot feed instances into the system, to perform instance-based analysis which is an effective strategy for understanding machine learning models.

Another way of handling large number of neurons is to employ  dimensionality reduction techniques.
By projecting a high-dimensional vector into two-dimensional space, we can better represent the high-dimensional nature of deep neural network models. 
Rauber et al.~\cite{rauber2017visualizing} studied how 2-D projected view of instance activations and neuron  filters can help users better understand neural network models.
Google's \textit{Embedding Projector}~\cite{smilkov2016embedding} tool, which is integrated into their Tensorflow deep learning framework~\cite{abadi2016tensorflow}, provides an interactive 3-D projection with some additional features (e.g., similar instance search).
ReVACNN~\cite{chung2016revacnn} is an interactive visual analytics system that uses dimensionality reduction for convolutional networks. 
While CNNVis~\cite{liu2017towards} uses clustering to handle large number of neurons, ReVACNN shows both   individual neurons and a 2-D projection embedded space (through t-SNE). 
The individual neuron view  helps users explore how individual neurons respond to a user-selected instance; 
the projected view can help them get a visual summary of instance activations.
However,  these two views  work independently.
It is difficult for users to combine their analyses, or compare multiple instances' neuron activations.

\section{Analytics Needs for Industry-Scale Problems}

The \name{} project started in April 2016.
Since its inception, we have conducted participatory design sessions with over 15 Facebook engineers, researchers, and data scientists across multiple teams to learn about their visual analytics needs.
Together, we collaboratively design and develop \name{} and  iteratively improve it.

In \autoref{sec:background}, we describe the workflow of how machine learning models are typically trained and used at Facebook, and how results are interpreted.
This discussion provides the background information and  context for which visualization tools may help improve deep learning model interpretation.

In \autoref{sec:requirements}, we summarize our main findings from our participatory design sessions 
to highlight six key design challenges that stem from Facebook's needs to work with large-scale datasets, complex deep learning model architectures, and diverse analytics needs.
These challenges have been inadequately addressed by current deep learning visualization tools,
and they motivate and shape our design goals for \name{}, which we will describe in \autoref{sec:goals}.

\subsection{Background:  Machine Learning Practice at Facebook}
\label{sec:background}

Facebook uses machine learning for some of their products. 
Researchers, engineers, and data scientists from  different teams at Facebook perform a wide range of machine learning tasks.

We first describe how Facebook's machine learning platform helps users train models and interpret their results.
Then, we present findings from our discussion with machine learning users and their common analytics patterns in interpreting machine learning models.
These findings guide our discovery of design challenges that \name{} aims to address.

\subsubsection{FBLearner Flow: Facebook's Machine Learning Platform}
\label{sec:flow}

To help engineers, including non-experts of machine learning, to more easily reuse algorithms in different products and manage experiments with ease, 
Facebook built a unified machine learning platform called \textit{FBLearner Flow}~\cite{flow,andrews2016productionizing}.
It supports many machine learning workflows.
Users can easily train models and see their results using the FBLearner Flow interface without writing any code.
For example, users can train a model by picking a relevant workflow from a collection of existing workflows and specifying several input parameters for the selected workflow
(e.g., location of training dataset, learning parameters).
The FBLearner Flow interface is particularly helpful for users who want to use existing machine learning models for their datasets without knowing their internal details.

Once the training process is done, the interface provides high-level information to aid result analysis (e.g., precision, accuracy).
To help users interpret the results from additional multiple aspects, several other statistics are available in the interface (e.g., partial dependence plots).
Users can inspect models' internal details via interactive visualization (e.g., for decision trees)~\cite{andrews2016productionizing}.
As deep neural network models gain popularity, 
developing visualization for their interpretation is a natural step for FBLearner Flow.

\subsubsection{Analytics Patterns for  Interpretation}
\label{sec:patterns}

To better understand how machine learning users at Facebook interpret model results, and how we may design \name{} to better support their analysis,
we conducted participatory design sessions with over 15 engineers and data scientists who regularly work with machine learning and deep neural network models.
At the high level, we learned that instance- and subset-based strategies are both common and effective, echoing findings from existing research.

\textbf{Instance-based analysis.}
One natural way for users at Facebook to understand complex models is by tracking how an individual example (i.e., training or test instance) behaves inside the models; 
users often have their own collection of example instances, for which they know their characteristics and ground truth labels.
Instance-level exploration is  especially useful when an instance is easy to interpret. 
For example, an instance consisting of text only is much easier to understand than an instance consisting of thousands of numerical features extracted from an end user's data.

\textbf{Subset-based analysis.}
Instance-based analysis, however, is insufficient for all cases.
Inspecting instances individually can be tedious, 
and sometimes hinder insight discovery, 
such as when instances are associated with many hard-to-interpret numerical features.
We learned that some Facebook researchers find subset-based  analysis to be more helpful for their work.
For example, suppose an instance represents an article that consists of many numerical features extracted from its attributes (e.g., length, popularity).
Some users would like to understand how the models behave at higher-level categorization (e.g., by topic, publication date). 
In addition, some users have curated instance subsets.
Understanding model behavior through such familiar subsets promotes their understanding.

\subsection{Design Challenges}
\label{sec:requirements}

Besides reaffirming the importance of two analysis strategies discussed above, and the need to support them simultaneously in \name{}, we have identified additional design challenges through the participatory design sessions.
We summarize them into six key design challenges. 
Thus far, they have not been adequately addressed by existing deep learning visualization tools.
And they shape the main design goals of \name{}, which we will describe in~\autoref{sec:goals}.

We have labeled the six challenges C1 -- C6 and have grouped them into three categories with the labels \textit{data}, \textit{model}, and \textit{analytics}, which indicate the causes for which the challenges arise.

\begin{enumerate}[label=C\arabic*., itemsep=1mm, topsep=2mm]

\item \textbf{Diverse input sources and formats} 
\tag{\sffamily \small \textsc{Data}}\\
While deep learning has become popular because of its superior performance for image  data,
it has also been applied to many different data formats, including text and numerical features~\cite{kim2014convolutional,joulin2016bag,deeptext,covington2016deep}. 
Furthermore, a single model may jointly use multiple types of data at a time.
For example, to classify a Facebook post, a model may jointly leverage its textual content, attached photos, and user information, each of which may be associated with many data attributes~\cite{deeptext}.
Working with such variety of data sources and formats opens up many opportunities for model interpretation; for example, we may be able to more easily categorize instances using their associated numerical features that can be more readily understood, instead of going the harder route of using image-based features. 

\item \textbf{High data volume} 
\tag{\sffamily \small \textsc{Data}}\\
Facebook, like many other companies, has a large amount of data.
The size of training data often reaches billions of rows and thousands of features.
This sheer size of data render many existing visualization tools unusable 
as they are often designed to visualize the whole dataset.

\item \textbf{Complex model architecture} 
\tag{\sffamily \small \textsc{Model}}\\
Many existing visualization tools for deep learning models often assume simple linear architectures where data linearly flow from the input layer to the output layer (e.g., a series of convolution and max-pooling layer in AlexNet)~\cite{yosinski2015understanding,liu2017towards,chung2016revacnn}.
However, most practical model architectures deployed in industry are very complex \cite{covington2016deep}; they are often deep and wide, consisting of many layers, neurons, and operations.

\item \textbf{A great variety of models} 
\tag{\sffamily \small \textsc{Model}}\\
Researchers and engineers at Facebook develop and evaluate models for  products every day.
It is important for visualization tools to be generalizable so they can  work with many different kinds of models.
A visualization system would likely be impractical to use or to deploy if a small change to a model requires significant changes made to existing code or special case handling.

\item \textbf{Diverse subset definitions} 
\tag{\sffamily \small \textsc{Analytics}}\\
When performing  subset-based analysis, users may want to define subsets in many different ways.
Since there are a large number of input formats and input features, there are numerous ways to specify subsets.
Instead of providing a fixed set of ways to define subsets, it is desirable to make this process flexible so that users can flexibly define subsets that are relevant to their tasks and goals.

\item \textbf{Simultaneous need for performing instance- and subset-level analysis} 
\tag{\sffamily \small \textsc{Analytics}}\\
Instance- and subset-based are complementary analytics strategies, and it is important to support both at the same time.
Instance-based analysis helps users track how an individual instance behaves in the models, but it is tedious to inspect many instances one by one.
By specifying subsets and enabling their comparison with individual instances, users can learn how the models respond to many different slices of the data.

\end{enumerate}

\section{\name{}: Visual Exploration of Neural Networks}
\label{sec:interface}

\begin{figure*}[t]
 \centering
 \includegraphics[width=\linewidth]{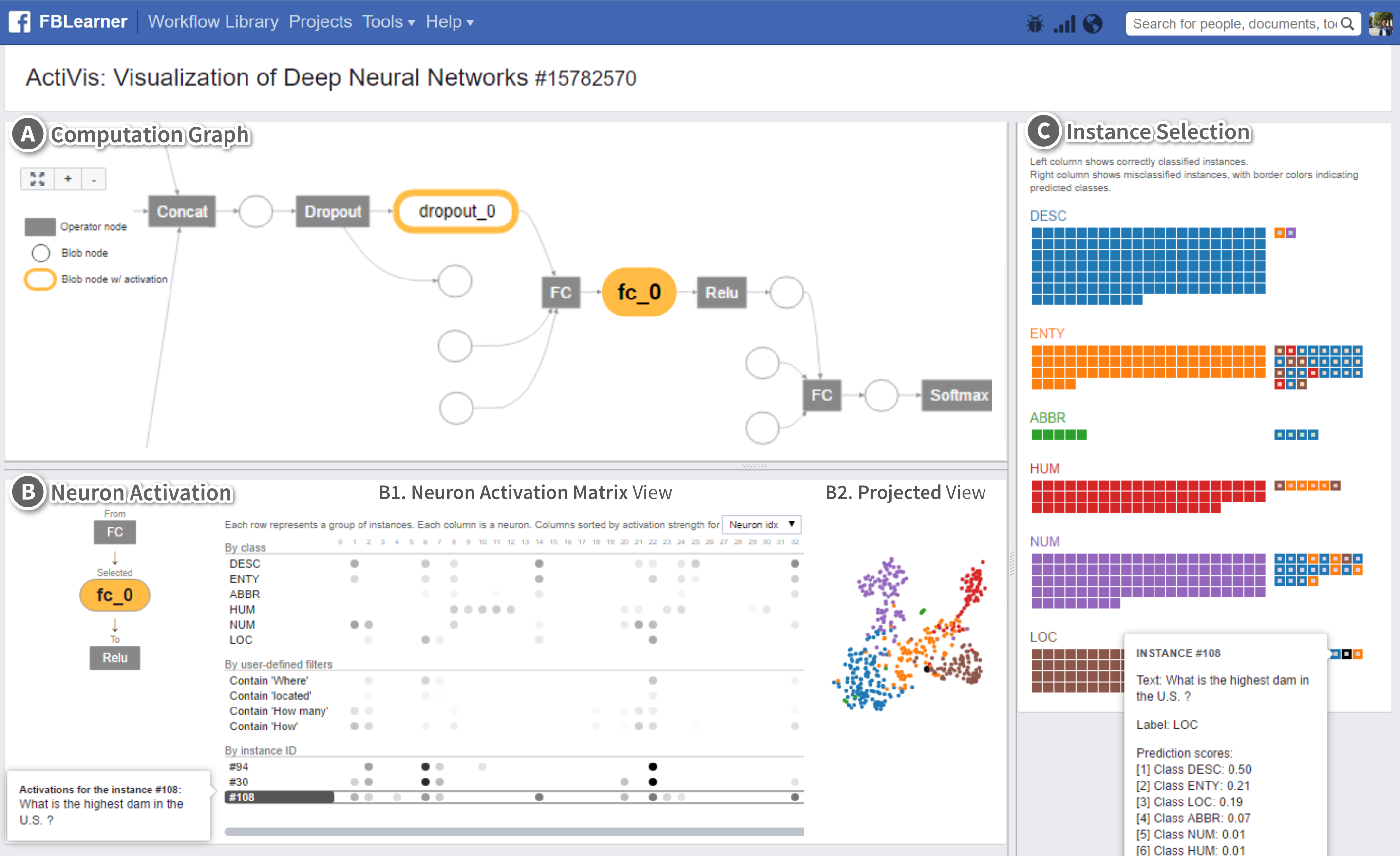}
 \caption{
 \name{} integrates multiple coordinated views.
\textbf{A.} The \textit{computation graph} summarizes the model architecture. 
\textbf{B.} The \textit{neuron activation} panel's \textit{matrix view} displays activations for instances, subsets, and classes (at B1), and its \textit{projected view} shows a 2-D t-SNE projection of the instance activations (at B2). 
\textbf{C.} 
The \textit{instance selection} panel displays instances and their classification results; correctly classified instances shown on the left, misclassified on the right.
Clicking an instance adds it to the neuron activation matrix view.
The dataset used is from the public TREC question answering data collections~\cite{li2002learning}.
The trained model is a word-level convolutional model based on~\cite{kim2014convolutional}.
}
 \label{fig:screenshot}
\end{figure*}

Through the design challenges we identified (in \autoref{sec:requirements}) in our participatory design sessions with researchers, engineers, and data scientists at Facebook,
we design and develop \name{}, a novel interactive visual tool for exploring a wide range of industry-scale deep neural network models.
In this section, we first present three main design goals distilled from our conversations with Facebook participants (\autoref{sec:goals}).
Then, for each design goal, we elaborate on how \name{} achieves it through its system design and visual exploration features (Sects. \ref{sec:main1}-\ref{sec:main3}).
We label the three design goals G1 -- G3.

\subsection{Design Goals}
\label{sec:goals}

\begin{enumerate}[label=G\arabic*.,itemsep=1mm, topsep=2mm]

\item \textbf{Unifying instance- and subset-based analysis to facilitate comparison of multiple instance activations.}
From our participatory design sessions, we learned that both instance- and subset-based analysis are useful and complementary.
We aim to support subset-level exploration by enabling users to flexibly define instance subsets for different data types (C1, C5),
e.g., a set of documents that contain a specific word. 
Subset-based analysis also allows users to explore datasets at higher-level abstraction, scaling to billion-scale data or larger (C2).
Furthermore, we would like to unify instance- and subset-level inspections to facilitate comparison of multiple instances and groups of instances in a single view (C6).

\item \textbf{Tight integration of overview of model architecture and localized inspection of activations.}
Industry-scale deep neural network models are often very complex, consisting of many operations (C3).
Visualizing every detail and activation value for all intermediate layers can overwhelm users.
Therefore, we aim to present the architecture of the models as a starting point of exploration, and let users switch to the detailed inspection of activations.

\item \textbf{Scaling to industry-scale datasets and models through  flexible system design.}
For \name{} to work with many different large-scale models and datasets used in practice, 
it is important for the system to be flexible and scalable.
We aim to support as many different kinds of data types and classification models as what FBLearner currently does (e.g., image, text, numerical) (C1, C4). 
We would like to achieve this by developing a flexible, modularized system that allows developers to use \name{} for their models with simple API functions, while addressing visual and computational scalability challenges through a multipronged approach (C2, C3).

\end{enumerate}

\subsection{Exploring Neuron Activations by Instance Subsets}
\label{sec:main1}

Drawing inspiration from existing visualizations~\cite{yosinski2015understanding,harley2015isvc,convnetjs,liu2017towards},
\name{} supports the visualization for individual instances.
However, it is difficult for users to spot interesting patterns and insights if he can only visualize one instance at a time.
For example, consider a hidden layer consisting of 100 neurons.
The neuron activations for an instance is a 100-dimension vector consisting of 100 numerical values, where each element in the vector does not have any specific meaning.
Instead, if multiple vectors of activation values are presented together, 
the user may more readily derive meaning by comparing them.
For example, users may find that some dimensions may respond more strongly to certain instances, 
or some dimensions are negatively correlated with certain classes.

A challenge in supporting the comparison of multiple instances stems  from the sheer size of data instances; 
it is impossible to present activations for all instances.
To tackle this challenge, we enable users to define \textit{instance subsets}.
Then we compute the average activations for instances within the subsets. 
The vector of average activations for a subset can then be placed next to the vectors of other instances or subsets for comparison.

The \textit{neuron activation matrix}, shown at \autoref{fig:screenshot}B.1, illustrates this concept of comparing multiple instances and instance subsets,
using the TREC question classification dataset\footnote{\small \url{http://cogcomp.cs.illinois.edu/Data/QA/QC/}}~\cite{li2002learning}.
The dataset consists of 5,500 question sentences and each sentence is labeled by one of six categories (e.g., is a question asking about \textit{location}?).
\autoref{fig:screenshot}B shows the activations for the last hidden layer 
of the word-level CNN model~\cite{kim2014convolutional,wildml}.
Each row represents either an instance or a subset of instances.
For example, the first row represents a subset of instances whose true class is `DESC' (descriptions).
Each column represents a neuron.
Each cell (circle) is a neuron activation value for a subset.
A darker circle indicates stronger activation.
This matrix view exposes the hidden relationships between neurons and data.
For instance, a user may find out a certain neuron is highly activated by instances whose true class is `LOC'.

\textbf{Flexible subset definition.} 
In \name{}, users can flexibly define instance subsets. 
A subset can be specified using multiple properties of the instances, in many different ways.
Example properties include raw data attributes, labels, features, textual content, output scores, and predicted label.
Our datasets consist of instances with many features and a combination of different types of data. 
Flexible subset definition enables users to analyze models from different angles.
For example, for instances representing text documents,
the user may create a subset for documents that contains a specific phrase.
For instances containing numerical features, 
users can specify conditions, using operations similar to relational selections in databases (e.g., \texttt{age > 20}, \texttt{topic = 'sports'}).
By default, a subset is created for each class
(e.g., a subset for the `DESC' class).

\begin{figure}[tb]
 \centering
 \includegraphics[width=\columnwidth]{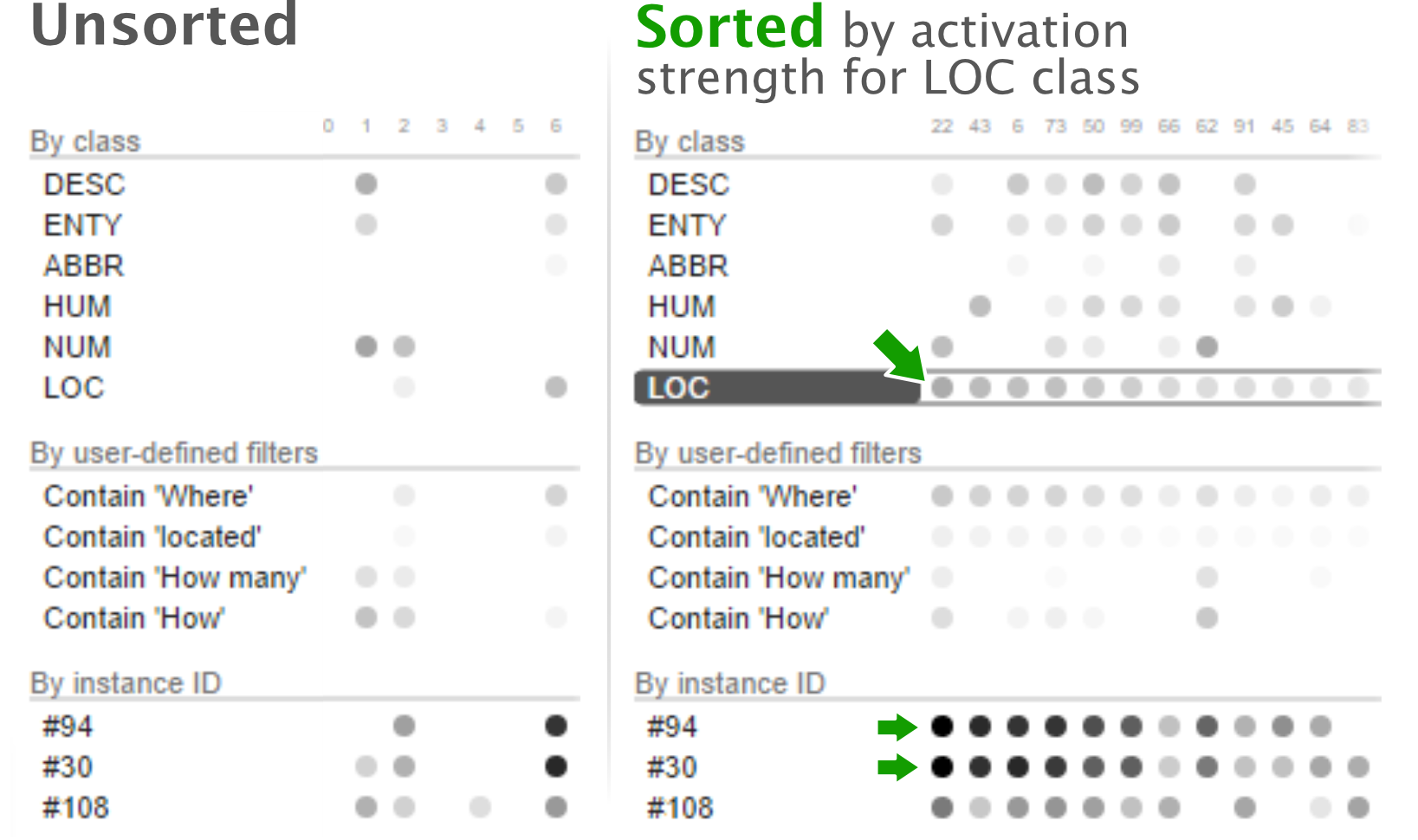}
 \caption{Sorting neurons (columns) by their average activation values for the \textit{LOC} (location) class helps users more easily spot instances whose activation patterns are positively correlated with that of the class, e.g., instances \#94 and \#30 (see green arrows).
 }
 \label{fig:sort}
\end{figure}

\textbf{Sorting to reveal patterns.}
The difficulty in recognizing patterns increases with the number of neurons.
\name{} allows users to sort neurons (i.e., columns) by their activation values. 
For example, in \autoref{fig:sort}, the neurons are sorted based on the average activation values for the class `LOC'.
Sorting facilitates activation comparison and helps reveal patterns, such as spotting instances that are positively correlated with their true class in terms of the activation pattern (e.g., instances \#94 and \#30 correlate with the `LOC' class in \autoref{fig:sort}).

\begin{figure}[tb]
 \centering
 \includegraphics[width=\columnwidth]{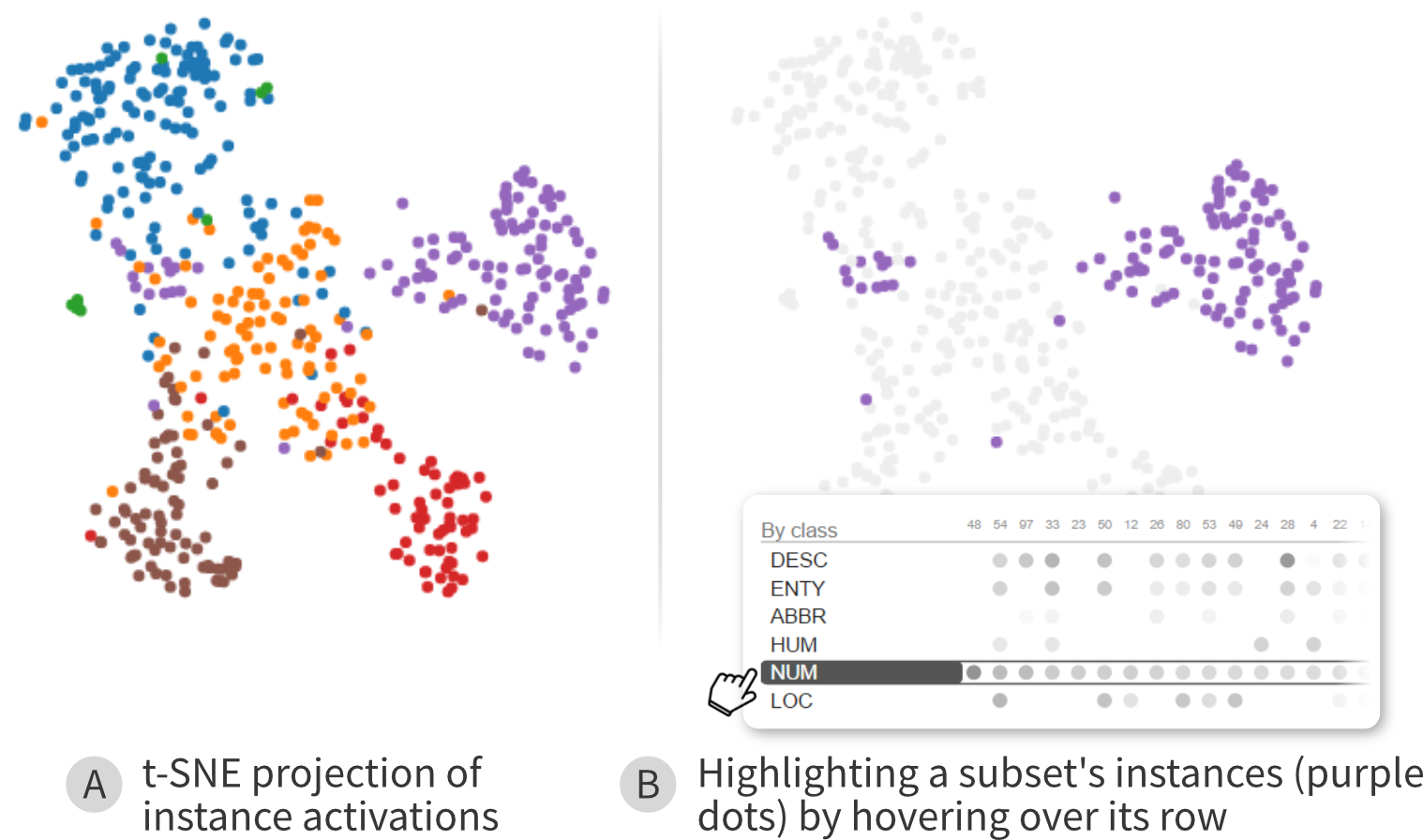}
 \caption{
Hovering over an instance subset (e.g., for the \textcolor{myPurple}{\textbf{\sffamily \small NUM}}ber class) highlights its instances (purple dots) in the t-SNE projected view.}
 \label{fig:projected}
\end{figure}

\textbf{2-D projection of activations.}
To help users visually examine instance subsets, \name{} provides a 2-D \textit{projected view} of instance activations.
Projection of high-dimensional data into 2-D space has been considered an effective exploration approach~\cite{rauber2017visualizing,smilkov2016embedding,chung2016revacnn,choo2010ivisclassifier}.
\name{} performs \textit{t-distributed stochastic neighbor embedding (t-SNE)}~\cite{maaten2008visualizing} of instance activations.
\autoref{fig:screenshot}B.2 shows an example where
each dot in the view represents an instance (colored by its true class), and instances with similar activation values are placed closer together by t-SNE.

The projected view complements with the \textit{neuron activation matrix} view (\autoref{fig:screenshot}B.1).
Hovering over a subset's row in the matrix would highlight the subset's instances in the projected view, allowing the user to see how instances within the subsets are distributed.
In the projected view, hovering over an instance would display its activations; clicking that instance will add it to the matrix view as a new row.

\subsection{Interface: Tight Integration of Model, Instances, and Activation Visualization}
\label{sec:main2}

The above visual representation of activations is the core of our visual analytics system. 
To help users interactively  specify where to start their exploration of a large model,
we designed and developed an integrated system interface.
As depicted in \autoref{fig:screenshot}, the interface consists of multiple panels.
We describe each of them below.

\subsubsection*{A: Overview of Model Architecture}

Deep learning models often consist of many operations, which makes it difficult for users to fully understand their structure.
We aim to provide an overview of the model architecture to users, so  they can first make sense of the models, before moving on to parts of the models that they are interested in.

Deep neural network models are often represented as computation graphs (DAGs) (as in many deep learning frameworks like Caffe2\footnote{\small \url{https://caffe2.ai/}}, TensorFlow~\cite{abadi2016tensorflow}, and Theano~\cite{bergstra2010theano}). 
The frameworks provide a set of operators (e.g., convolution, matrix multiplication, concatenation) to build machine learning programs, and model developers (who create new machine learning workflows for FBLearner Flow) write the programs using these building blocks.
Presenting this graph to users would help them first understand the structure of the models and find interesting layers to explore the detailed activations.

There are several possible ways in visualizing computation graphs.
One approach is to represent operators as nodes and variables as edges.
This approach has gained popularity, thanks to its adoption by TensorFlow.
Another way is to consider both an operator and a variable  as a single node. 
Then the graph becomes a bipartite graph: the direct neighbors of an operator node are always variable nodes; the neighbors of a variable node are always operator nodes.
Both approaches have their pros and cons.
While the first approach can have a compact representation by reducing the number of nodes,
the second one, a classical way to represent programs and diagrams, makes it easier to track data.
For  \name{}, it would be better to make variable nodes easy to locate as we present activations for a selected variable. Therefore,
we decided to represent the graph using the second approach.

The  visualization of the computation graph is shown on the top panel (\autoref{fig:screenshot}A).
The direction of data flow is from left (input) to right (output).
Each node represents either an operator (dark rectangle) or tensor (circle).
To explore this medium-sized graph (often $>$100 nodes), users can zoom and pan the graph using a mouse. 
When users hover over a node, its full name is shown, and
when they click it, its corresponding activation is shown in the neuron activation panel.

\subsubsection*{B: Activation for Selected Node}

When users select a node of interest from the computation graph,
the corresponding neuron activation panel (\autoref{fig:screenshot}B) will be added to the bottom of the computation graph panel.
The neuron activation panel has three subpanels: (0) the names of the selected node and its neighbors, (1) the neuron activation matrix view, and (2) the projected view.
The left subpanel shows the name of the selected variable node and its neighbors.
Users can hover over a node to highlight where it is located in the computation graph on the top.
The neuron matrix view (\autoref{fig:screenshot}B.1) and projected view (\autoref{fig:screenshot}B.2) show instance activations for the selected node.
Note that we described these views in \autoref{sec:main1}.

\begin{figure}[tb]
 \centering
 \includegraphics[width=\columnwidth]{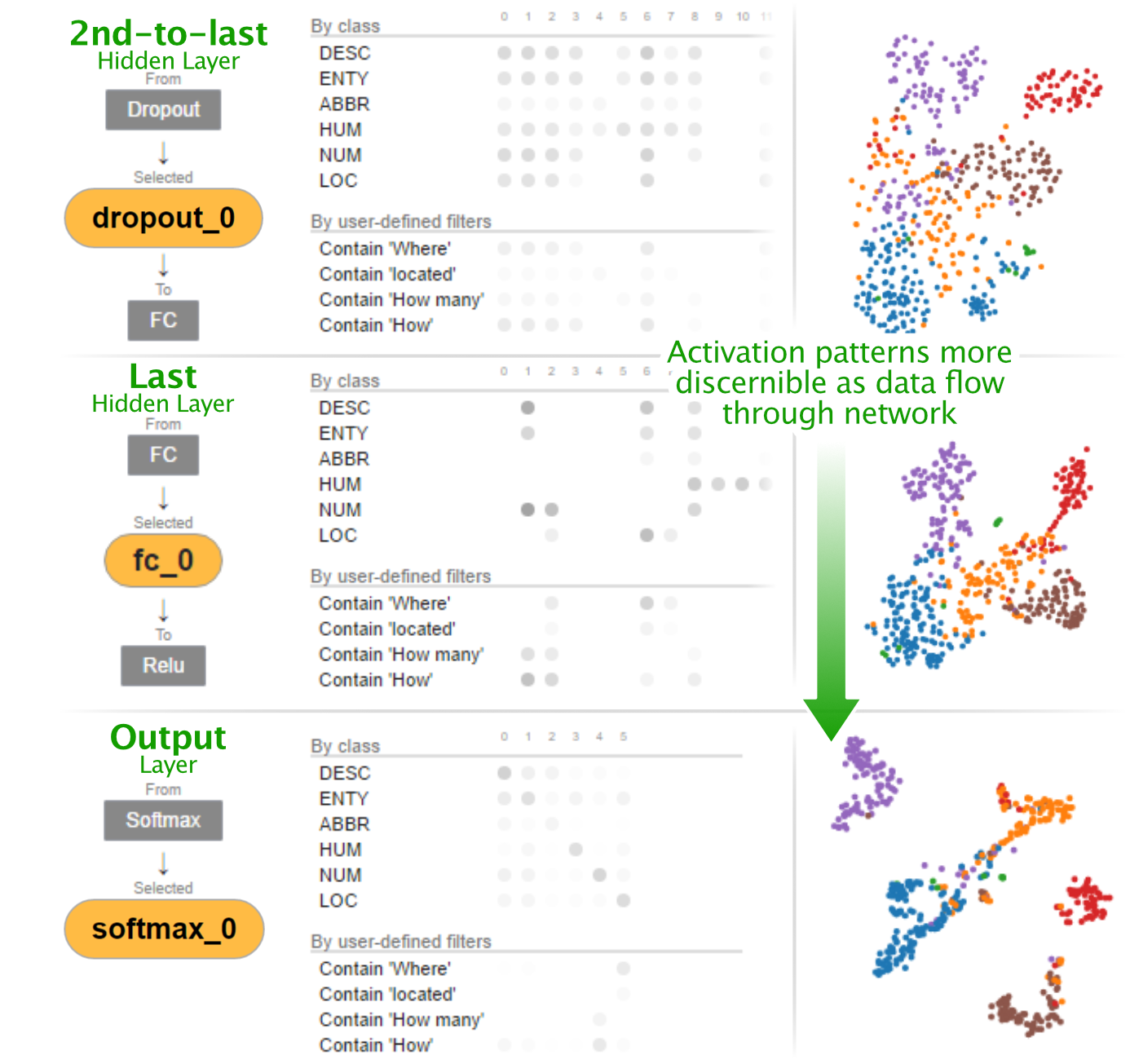}
 \caption{Users can simultaneously visualize and compare multiple layers' activations. 
Shown here, from top to bottom, are: the second-to-last hidden layer, the last hidden layer, and the output layer.
Their projected views show that as instances flow through the network from input (top) to output (bottom),
their activation patterns  gradually become more discernible and clustered (in projected view).
}
 \label{fig:multiple}
\end{figure}

Users can select multiple nodes and visually compare their activation patterns.
\autoref{fig:multiple} illustrates that users can visually explore how models learned   the hidden structure of data through multiple layers.
The figure shows three layers, from top to bottom: the second-to-last hidden layer which concatenates multiple maxpool layers~\cite{kim2014convolutional}, the last hidden layer, and the output layer.  
As shown in the figure, the layer's projected views show that as data flow through the network, from input (top) to output (bottom), neuron activation patterns gradually become more discernible and clustered.

\subsubsection*{C: Instance Selection}

The instance selection panel helps users get an overview of instances with their prediction results and determine which ones should be added to the neuron activation view for further exploration and comparison.

The panel is located at the right side on the interface.
It visually summarizes prediction results.
Each square represents an instance.
Instances are vertically grouped based on their true label.
Within a true label (row group), 
the left column shows correctly classified instances, sorted by their prediction scores in descending order (from top to bottom, and left to right within each row).
The right column shows misclassified instances.
An instance's fill color represents its true label, its border color the predicted label.
When the user hovers over an instance, a tooltip will display  basic information about the instance  (e.g., textual content, prediction scores).

The panel also helps users determine which instances can be added to the activation view for further exploration.
By hovering over one of the instance boxes,
users can see the instance's activations. 
A new row is added to the activation  view presenting the activation values for the selected instance.
When users' mouse leaves the box, the added row disappears. To make a row persistent, users can simply click the box.
In a similar fashion, users can add many rows by clicking the instance boxes.
Then, they can compare activations for multiple instances and also compare those for instances with those for groups of instances.

\subsection{Deploying \name{}: 
Scaling to Industry-scale Datasets and Models}
\label{sec:main3}

We have deployed \name{} on  Facebook's machine learning platform.
Developers who want to use \name{} for their model can easily do so by adding only a few lines of code, which instructs their models' training process to generate information needed for \name{}'s visualization.
Once model training has completed,
the FBLearner Flow interface provides the user with a link to \name{} to visualize and explore the model.
The link opens in a new web browser window.

\name{} is designed to work with classification tasks that use deep neural network models.
As complex models and large datasets are commonly used at Facebook, it is important that \name{} be scalable and flexible, so that engineers can easily adopt \name{} for their models.
This section describes our approaches to building and deploying \name{} on FBLearner, Facebook's machine learning platform.

\subsubsection{Generalizing to Different Models and Data Types}
One of our main goals is to support as many different kinds of data types and models as what FBLearner currently does (e.g., images, text, numerical).
The key challenge is to enable existing deployed models to generate data needed for \name{} with as little modification as possible.
Without careful thinking, we would have to add a large amount of model-specific code, to enable \name{} to work with different models.
To tackle this challenge, we modularize the data generation process  and define API functions for model developers so that they can simply call them in their code, to activate \name{} for their models. 
In practice, for a developer to use \name{} for a model, only three function calls are needed to  be added (i.e., calling the \textit{preprocess}, \textit{process}, and \textit{postprocess} methods).
For example, developers can specify a list of variable nodes that users can explore, as an argument of the \textit{preprocess} function (described in detail in \autoref{sec:scale}).
Furthermore, developers can leverage \textit{user-defined functions} to specify how subsets are defined in \name{}, a capability particularly helpful for the more abstract, unstructured data types, such as image and audio.
For example, developers may leverage the output of an object recognition algorithm that detects objects (e.g., cats, dogs) to define image subsets (e.g., subset of images that contain dogs).

\subsubsection{Scaling to Large Data and Models}
\label{sec:scale}

\name{} addresses visual and computational scalability challenges through multiple complementary approaches. 
Some of them were introduced in earlier sections (e.g., \autoref{sec:main1}), 
such as \name{}'s overarching subset-based analysis,
and the simultaneous use of \textit{neuron matrix} (for individual neuron inspection) and \textit{projected view} (in case of many neurons).
We elaborate on some of our other key ideas below.

\textbf{Selective precomputation for variable nodes of interest.}
Industry-scale models often consist of a large number operations (i.e., variable nodes), up to hundreds.
Although any variable node can be visualized in the activation visualization,
if we compute activations for all of them,
it will require significant computation time and  space for storing the data.
We learned from our discussion with experts and design sessions with potential users  that it is typical for only a few variable nodes in a model to be of particular interest (e.g., last hidden layer in CNN).
Therefore, instead of generating activations for all variable nodes, we let model developers specify their own default set of variable nodes.
The model developers can simply specify them as an argument of the \textit{preprocess} method.
To explore variable nodes not included in the default set, a user can add them by specifying the variable nodes in the FBLearner Flow interface. 
Such nodes will then be available in the computation graph (highlighted in yellow).

\textbf{User-guided sampling and visual instance selection.}
For billion-scale datasets, it is undesirable to display all data points in the instance selection panel.
Furthermore, we learned from our design sessions that researchers and engineers are primarily interested in a small number of representative examples, such as ``test cases'' that they have curated 
(e.g.,  instances that should be labeled as Class `LOC' by all well-performing models).
To meet such needs, by default, we present a sample of instances  in the interface (around 1,000), 
which meet the practical needs of most Facebook engineers.
In addition, users may also guide the sampling to include arbitrary examples that they specify (e.g., their test cases).

\textbf{Computing neuron activation matrix for large datasets.}
The main computational challenge of \name{} is in computing the neuron activation matrix over large datasets.
Here, we describe our scalable approach whose time complexity is linear in the number of data instances. 
We first create a matrix $S$ (\#instances $\times$ \#subsets) that describes all instance-to-subset mappings.
Once a model predicts labels for instances, 
it produces an activation matrix $A$ (\#instances $\times$ \#neurons) for each variable node.
By multiplying these two matrices (i.e., $S^T A$), followed by normalization, we obtain a matrix containing all subsets' average neuron activation values, which are visualized in the neuron matrix view.
As the number of instances dominates, the above computation's time complexity is linear in the number of instances. 
In practice, this computation roughly takes the same amount of time as testing a model.
We have tested \name{} with many datasets (e.g., one with 5 million training instances).
\name{} can now scale to any data sizes that FBLearner supports (e.g., billion-scale or larger).

\subsubsection{Implementation Details}

The visualization and interactions are implemented mainly with React.js.\footnote{\small \url{https://facebook.github.io/react/}}
We additionally use a few D3.js V4 components.\footnote{\small \url{https://d3js.org/}}
The computation graph is visualized using Dagre,\footnote{\small \url {https://github.com/cpettitt/dagre}} a JavaScript library for rendering directed graphs.
All the backend  code is implemented in Python (including scikit-learn\footnote{\small \url {http://scikit-learn.org/}} for t-SNE) and the activation data generated from backend are passed to the interface using the JSON  format.

\section{Informed Design through Iterations}

The current design of \name{} is the result of twelve months of investigation
and development effort through many iterations.

\begin{figure}[t]
 \centering
 \includegraphics[width=0.9\columnwidth,trim={0cm 5.3cm 0cm 0cm},clip]{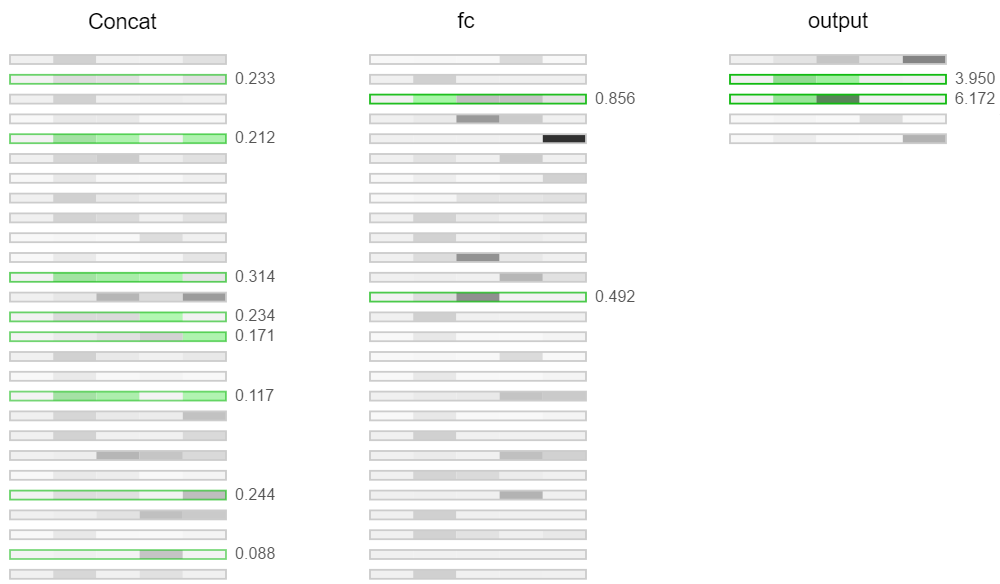}
 \vspace{-5pt}
 \caption{
 Version 1 of \name{},
 showing an instance's neuron activation strengths, encoded using color intensity.
 A main drawback of this design was that users could only see the activations for a single instance at a time.
 Activation comparison across multiple instances was not possible.
 }
 \label{fig:early1}
\end{figure}

\begin{figure}[tb]
 \centering
 \includegraphics[width=0.95\columnwidth]{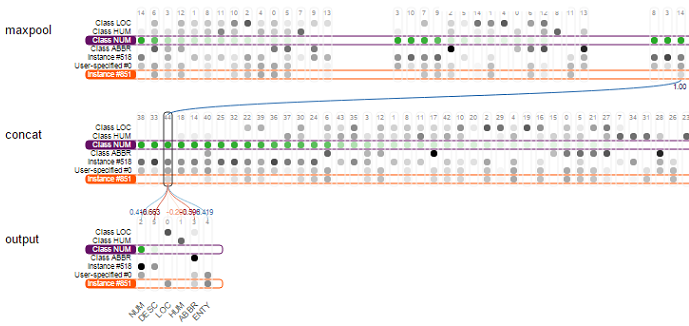}
 \vspace{-10pt}
 \caption{Version 2 of \name{},
 which unified instance- and subset-level activation visualization.
 This design was too visually overwhelming and did not scale to complex models,
 as it allocated a matrix block for each operator;
 a complex model could have close to a hundred operators.
 }
 \label{fig:early2}
\end{figure}

\textbf{Unifying instances and subsets to facilitate comparison of multiple instances.}
The first version of \name{}, depicted in \autoref{fig:early1}, visualizes activations for all layers
(each column group represents  a single layer).
A main drawback of this design is that users can only see the activations for a single instance at a time; they cannot compare  multiple instances' activations.
While, for the subsets, we use an approach similar to \name{}'s design (each dot represents the average values for the subset),
we encode activations for a given instance using background color (here, in green). This means that the visualization cannot support activation comparison across multiple instances.
This finding prompted us to  unify the treatment for instances and subsets to enable comparison across them.
\autoref{fig:early2} shows our next design iteration that implements this idea.

\textbf{Separating program and data to handle complex models.}
Although the updated version (\autoref{fig:early2}) shows activations for multiple instances, which helps users explore more information at once, it becomes visually too overwhelming when visualizing large, complex models.
Some engineers expressed concern that this design might not generalize well to different models.
Also, engineers are often interested in only a few variable nodes, rather than  looking at many variable nodes.
Therefore, we decided to separate the visualization of the model architecture and the activations for a specific variable node.

\textbf{Presenting 2-D projection of instances.}
One researcher suggested that \name{} should provide more detail for  each neuron, in addition to \textit{average} activations.
Our first solution was to present statistics (e.g., variance) and distributions for each neuron.
However, some researchers cautioned that this approach could be misleading, 
because these summaries might not fully capture high-dimensional activation patterns.
This prompted us to add the projected view (t-SNE), which enabled users to better explore the high-dimensional patterns (see \autoref{fig:projected}).

\section{Case Studies \& Usage Scenarios}
\label{sec:evaluations}

To better understand how \name{} may help Facebook machine learning users with their interpretation of deep neural network models, we recruited three Facebook engineers and data scientists to use the latest version of \name{} to explore text classification models relevant to their work. 
We summarize key observations from these studies to highlight \name{}'s benefits (\autoref{sec:cases}).
Then, based on observations and feedback from these users and others who participated in our earlier participatory design sessions, we present example usage scenarios for ranking models to illustrate how \name{} would generalize (\autoref{sec:ranking}).

\subsection{Case Studies: Exploring Text Classification Models with \name{}}
\label{sec:cases}

\subsubsection{Participants and Study Protocol}

We recruited three Facebook engineers and data scientists to use our tools (their names substituted for privacy):

\begin{description}[itemindent=0mm, leftmargin=3mm, topsep=1mm, itemsep=1mm]

\item \textit{Bob} is a software engineer who has expertise in natural language processing. 
He is experimenting with applying text classification models to some Facebook experiences, such as for detecting intents from a text snippet, like understanding when the user may want to go somewhere~\cite{deeptext}.
For example, suppose a user writes \textit{``I need a ride''}, Bob may want the models to discover if the user needs transportation to reach the destination.
He is interested in selecting the best models based on experimenting with many parameters and a few different models, as in~\cite{joulin2016bag,kim2014convolutional}.
\item \textit{Dave} is a relatively new software engineer.
Like Bob, he is also working with text classification models for user intent detection, but unlike Bob, he is more interested in preparing training datasets from  large collections of databases.
\item \textit{Carol} is a data scientist who holds a Ph.D. in the area of natural language processing. 
Unlike Bob and Dave, she is working with many different machine learning tasks, focusing on textual data.
\end{description}

We had a 60-minute session with each of the three participants.
For the first 20 minutes, we asked them a few questions about their typical workflows, and how they train models and interpret  results.
Then we introduced them to \name{} by describing its  components. 
The participants used their own datasets and models, available from FBLearner Flow.
After the introduction, the participants used \name{} while thinking aloud.
They also gave us feedback on how we could further improve \name{}.
We recorded audio during the entire session and video for the last part.

\subsubsection{Key Observations}

We summarize our key observations from interacting with the three participants into the following three themes, each highlighting how our tool helped them with the analysis.

\textbf{Spot-checking models with user-defined instances and subsets.}
\name{} supports flexible subset definition.
This feature was developed based on the common model development pattern where practitioners often curate ``test cases'' that they are familiar with, and for which they know their associated labels.
For example, a text snippet ``Let's take a cab'' should be classified as a positive class of detecting transportation-related intent.
Both  Bob and Dave indeed found this feature useful (i.e., they also had their own ``test cases''), and they appreciated the ability to specify and use their own cases.
This would help them better understand whether their models are working well, by comparing the activation patterns of their own instances with those of other instances in the positive or negative classes.
Bob's usage of \name{} and comments echo and support the need for subset-level visualization and exploration, currently inadequately supported by existing tools.

\textbf{Graph overview as a crucial entry point to model exploration.}
From our early participatory design sessions, we learned that \name{}'s graph overview was  important for practitioners who work with complex models 
whose tasks only require them to focus on specific components of the models.
Bob, who works with many different variations of text classification models, 
has known that the model he works with  mainly uses convolution operations and  was curious to see how the convolution works in detail.
When he launched \name{}, he first examined the model architecture around the convolution operators using the computation graph panel. 
He appreciated that he could see how model training parameters are used in the model, which helped him develop better understanding of the  internal working mechanism of the models.
For example, he found how and where \textit{padding} are used in the models by exploring the graph~\cite{wildml}.
After he got a better sense about how the model function around the convolution operators, he examined the activation patterns of the convolution output layer.
This example shows that the graph overview is important for understanding complex architectures and locating parts that are relevant to the user's tasks.
In other words, the graph serves as an important entry point of Bob's analysis.
Existing tools assuming user familiarity with models may not hold 
in  real-world large-scale deployment scenarios.

\textbf{Visual exploration of activation patterns for evaluating model performances and for debugging hints.}
One of the main components of \name{} is the visual representation of activations  that helps users easily recognize patterns and anomalies.
As Carol interacted with the visualization, she gleaned a number of new insights, and a few hints for how to debug deep learning models in general.
She interactively selected many different instances  and added them to the neuron activation matrix to see how they activated neurons.
She found out that the activation patterns for some instances are unexpectedly similar, even though the textual content of the instances seem very different.
Also, she spotted that some neurons were not activated at all.
She hypothesized that the model could be further improved by changing some of the training parameters, so
she decided to modify them to improve the  model.
While the neuron activation panel helps Carol find models that can be further improved,
Bob found some interesting patterns from the activation patterns for the convolution output layer.
He quickly found out that some particular words are highly activated while some other words, which he thought can be highly activated, do not respond much.
This helped him identify  words that are potentially more effective for classification.
The examples above demonstrate the power of visual exploration. \name{} helps users recognize patterns by interacting with instances and instance subsets they are familiar with.

\subsection{Usage Scenario: Exploring  Ranking Models}
\label{sec:ranking}

As there are many potential uses for \name{} at Facebook, we also discussed with a number of researchers and engineers at different teams to understand how they may adopt \name{}.
Below, we present a usage scenario of \name{} for exploring ranking models, based on our discussion.
We note the scenario strongly resembles others that we have discussed so far;
this is encouraging because enabling \name{} to generalize across teams and models is one of our main goals.

Alice is a research scientist working with ranking models,
one of the important machine learning tasks in industry.
The ranking models can be used to recommend relevant content 
to users by analyzing a large number of numerical features extracted from databases~\cite{backstrom2016serving,he2014practical}.
Alice is experimenting with deep neural network models to evaluate how these models work for a number of ranking tasks.
She often performs subset-based analysis when examining model performance, such as defining subsets based on categories of page content.
Subset-based analysis is essential for Alice, 
because she works with very large amount of training data (billions of data points, thousands of features). 
\name{}'s instance-based exploration feature is not yet helpful for Alice, since she is still familiarizing herself with the data and has not identified instances that she would like to use for spot-checking the model.
In \name{}, Alice is free to use either or both of instance- and subset-based exploration.
For new, unfamiliar datasets, Alice finds it much easier to start her analysis from the high level, then drill down into subsets, using attributes or features.

Alice has trained a fully-connected deep neural network model with some default parameters.
When she launches \name{}, she first examines the output layer to see how the  activation patterns for the positive and negative classes may be different.
To her surprise, they look similar.
Furthermore, by inspecting the neuron activation matrix view, she realizes that many neurons are not activated at all --- their activation values are close to 0.
This signals that the model may be using more neurons than necessary.
So, she decided to train additional models with  different parameter combinations (e.g., reduce neurons) to relieve the above issue.

The performances of some models indeed improve.
Happy with this improvement, Alice moves on to perform deeper analysis of the trained models.
She first creates a number of  instance subsets by using  \textit{features}. 
She utilizes  50 top features known to be important for ranking.
For categorical features, 
she defines a subset for each category value.
For numerical features, she quantizes them into a small number of subsets based on the feature value distribution.
\name{}'s neuron activation matrix view visualizes how the subsets that Alice has defined are activating the neurons.
Maximizing the matrix view to take up the entire screen (and minimizing the computation graph view), 
Alice visually explores the activation matrix and identifies 
a number of informative, distinguishing activation patterns.
For example, one neuron is highly activated for a single subset, and much less so for other subsets, suggesting that neuron's potential predictive power.
With \name{}, Alice can train models that perform well and understand how the models capture the structure of datasets by examining the relationships between features and neurons.

\section{Discussion and Future Work}
\label{sec:future}

\indent\indent
\textbf{Visualizing gradients.} 
Examining \textit{gradients} is one of the effective ways to explore deep learning models~\cite{convnetjs,chung2016revacnn}. It is straightforward to extend \name{} to visualize gradients by replacing activations with gradients. While activation represents forward data flow from input to output layers, gradient represents backward flow.
Gradients would help developers to locate neurons or datasets where the models do not perform well.

\textbf{Real-time subset definition.}
For \name{} to work with a new subset, it needs to load the dataset into RAM to check which instances satisfy the subset's conditions.
Currently, it is not of high priority for the above process to be performed in real time, 
because users often have pre-determined subsets to explore.
We plan to integrate dynamic filtering and searching capabilities, to speed up both subset definition and instance selection.

\textbf{Automatic discovery of interesting subsets.}
With \name{}, users can flexibly specify subsets in infinitely many ways.
One of the engineers  commented that \name{} could help suggest interesting subsets for exploration, based on heuristics or measures.
For example, for text datasets, such a subset could include phrases whose activation patterns are very similar or different to those for a given instance or class.

\textbf{Supporting input-dependent models.}
An interesting research direction is to extend \name{} to support models that contain variable nodes whose number of neurons changes depending on the input (e.g., the number of words in a document), and to study the relationships between neurons and subsets for such cases.

\textbf{Understanding how \name{} informs model training.}
We plan to conduct a longitudinal study to better understand  \name{}'s impact on Facebook's machine learning workflows, such as how \name{} may inform the model training process. 
For example, a sparse neuron matrix may indicate that a model is using more neurons than needed, which could inform engineers on their decisions for hyperparameter tuning.

\section{Conclusion}
\label{sec:conclusion}

We presented \name{}, a visual analytics system for deep neural network models.
We conducted participatory design session with over 15
researchers and engineers across many teams at Facebook to 
identify  key design challenges,
and based on them, we distilled three main design goals: (1) unifying instance- and subset-level exploration;
(2) tight integration of model architecture and localized  activation inspection; and (3) scaling to industry-scale data and models.
\name{} has been deployed on Facebook's machine learning platform.
We presented case studies with Facebook engineers and data scientists, and usage scenarios of how \name{} may be used with different applications.


\acknowledgments{
We thank Facebook Applied Machine Learning Group, especially Yangqing Jia, Andrew Tulloch, Liang Xiong, and Zhao Tan for their advice and feedback.
This work is partly supported by the NSF Graduate Research Fellowship Program under Grant No. DGE-1650044.}

\clearpage

\bibliographystyle{abbrv}

\bibliography{references}
\end{document}